\documentclass[notitlepage,aps,12pt,eqsecnum,tightenlines]{revtex4}

\usepackage{amsmath}
\usepackage{amssymb,amsfonts}
\usepackage{bm}
\usepackage[mathcal]{euscript}
\usepackage{graphicx}
\usepackage{psfrag}
\usepackage{subfigure}

\newcommand{\comment}[1]{}

\newcommand{\ie}{\textit{i.e.}}
\newcommand{\etal}{\textit{et~al.}}

\newcommand{\mathnotation}[2]{\newcommand{#1}{\ensuremath{#2}}}

\renewcommand{\l}{\left}			
\renewcommand{\r}{\right}			
\mathnotation{\pd}{\partial}			
\mathnotation{\ee}{{\mathrm e}}			
\mathnotation{\imi}{\mathrm{i}}			
\mathnotation{\ldef}{\mathrel{\raisebox{.069ex}{:}\!\!=}}
\mathnotation{\rdef}{\mathrel{=\!\!\raisebox{.069ex}{:}}}
\mathnotation{\dint}{\,{\mathrm{d}}}		

\mathnotation{\grad}{\nabla}			
\renewcommand{\div}{\grad\cdot}			
\mathnotation{\curl}{\grad\times}		
\mathnotation{\lapl}{\nabla^2}			

\renewcommand{\time}{t}				
\mathnotation{\iter}{n}				
\mathnotation{\iterj}{m}			
\mathnotation{\xc}{x}				
\mathnotation{\xv}{{\bm{\xc}}}			
\mathnotation{\Xc}{X}				
\mathnotation{\Xv}{{\bm{\Xc}}}			
\mathnotation{\velc}{v}				
\mathnotation{\velv}{{\bm{\velc}}}		
\mathnotation{\Velc}{V}				
\mathnotation{\Velv}{{\bm{\Velc}}}		

\mathnotation{\eps}{K}				
\mathnotation{\M}{\mathcal{M}}			
\mathnotation{\Mc}{M}				
\mathnotation{\ML}{\mathbb{\Mc}}		
\mathnotation{\twotorus}{{\mathcal{T}^2}}	

\mathnotation{\kv}{\bm{k}}
\mathnotation{\lv}{\bm{\ell}}
\mathnotation{\G}{\mathcal{G}}
\mathnotation{\Time}{T}
\mathnotation{\A}{A}
\mathnotation{\B}{B}
\mathnotation{\E}{E}

\mathnotation{\Diffc}{\kappa}			
\mathnotation{\Difft}{\mathbb{D}}		
\mathnotation{\scaledDiffc}{\epsilon}		

\mathnotation{\edir}{e}				
\mathnotation{\edirv}{\mathbf{\edir}}		
\mathnotation{\ediru}{\hat{\edir}}		
\mathnotation{\ediruv}{\hat{\mathbf{\edir}}}	
\mathnotation{\ediruinf}{\ediru^\infty}		
\mathnotation{\ediruvinf}{\ediruv^\infty}	
\mathnotation{\edirt}{\tilde{\edir}}
\mathnotation{\edirtv}{\tilde{\mathbf{\edirv}}}
\mathnotation{\sdir}{{\mathrm{s}}}		
\mathnotation{\sdirv}{\mathbf{s}}		
\mathnotation{\sdiru}{\hat{\sdir}}		
\mathnotation{\sdiruv}{\hat{\sdirv}}		
\mathnotation{\sdiruinf}{\sdiru^\infty}		
\mathnotation{\sdiruvinf}{\sdiruv^\infty}	
\mathnotation{\udir}{{\mathrm{u}}}		
\mathnotation{\udirv}{\mathbf{u}}		
\mathnotation{\udiru}{\hat{\udir}}		
\mathnotation{\udiruv}{\hat{\udirv}}		
\mathnotation{\udiruinf}{\udiru^\infty}		
\mathnotation{\udiruvinf}{\udiruv^\infty}	

\mathnotation{\metric}{g}			
\mathnotation{\gpert}{h}			
\mathnotation{\nugr}{\Lambda}			

\mathnotation{\uvx}{\ediruv_1}

\newcommand{\adeq}{advection--diffusion equation}

\begin{document}

\title{The Strange Eigenmode in Lagrangian Coordinates}
\author{Jean-Luc Thiffeault}
\date{\today}
\email{jeanluc@imperial.ac.uk}
\affiliation{Department of Mathematics, Imperial College London,
SW7 2AZ, United Kingdom}

\begin{abstract}
For a distribution advected by a simple chaotic map with diffusion, the
``strange eigenmode'' is investigated from the Lagrangian (material) viewpoint
and compared to its Eulerian (spatial) counterpart.  The eigenmode embodies
the balance between diffusion and exponential stretching by a chaotic flow.
It is not strictly an eigenmode in Lagrangian coordinates, because its
spectrum is rescaled exponentially rapidly.
\end{abstract}


\maketitle

\textbf{There are two main types of coordinates used to represent fluid flow
and dynamical systems.  Eulerian (or spatial) coordinates are fixed in space,
while Lagrangian (or material) coordinates follow parcels of fluid.  Strange
eigenmodes are persistent patterns in mixing---they can decay slowly, and
hence remain visible in the concentration field for a long time.  So far,
these have been studied from the Eulerian viewpoint.  Here we describe the
nature of the strange eigenmode in Lagrangian coordinates for a simple map. It
is not a true eigenmode because its wavelength is continuously rescaled in
time.}

\section{Introduction}

The enhanced mixing of a passive scalar is one of the most direct consequences
of chaos: a flow whose trajectories exhibit sensitivity to initial conditions
will lead to rapid mixing.  There are powerful theories based on the
distribution of Lyapunov exponents~\cite{Antonsen1996,Balkovsky1999,Son1999}
that link the mixing rate of the passive scalar with the chaotic properties of
the flow.  It has been recently suggested, following earlier work of
Pierrehumbert~\cite{Pierrehumbert1994,Fereday2002,Wonhas2002,Sukhatme2002,%
Thiffeault2003d,Pikovsky2003,Liu2004}, that the mixing properties of the flow
can often be elucidated only by solving a full eigenvalue problem for the
advection--diffusion operator, in an analogous manner to what is done for the
kinematic dynamo~\cite{STF}.  The resulting eigenfunctions have been dubbed
\emph{strange eigenmodes} by Pierrehumbert, and are closely related to
Pollicott--Ruelle resonances in ergodic
theory~\cite{Pollicott1981,Pollicott1986,Ruelle1986}, which describe the
long-time decay of correlations in mixing hyperbolic dynamical systems.
Strange eigenmodes have also been observed
experimentally~\cite{Rothstein1999,Voth2003}.  They are often called
\emph{persistent patterns} or \emph{large-scale eigenfunctions}.

The strange eigenmodes reflect a balance between advection and diffusion.  On
its own, advection is incapable of achieving mixing: it shuffles the
concentration field of the passive scalar but does not decrease its
fluctuations.  The role of advection is to stir the concentration field,
thereby creating sharp gradients in concentration.  Physically, these
gradients are reflected in the filamentation experienced by a blob of dye when
it is stirred.  The sharp gradients enhance the role of diffusion
tremendously, and this allows mixing to proceed.  As the diffusivity of the
scalar is made smaller, the scale at which this mixing occurs decreases, so
the concentration field appears very rough.  In the limit of arbitrarily small
diffusivity, the concentration field is not smooth: it consists of a
superposition of strange eigenmodes.  The dominant one among these
eigenfunctions is called \emph{the} strange eigenmode, although there may be
several of comparable importance.

In the present work we tie the two types of theories together (\ie, Lyapunov
exponent-based and strange eigenmode) for a specific system, which was already
studied in~\cite{Thiffeault2003d} from the strange eigenmode viewpoint.  The
strange eigenmode represents a fundamentally Eulerian (spatial) view of
mixing, whereas the Lyapunov exponent view is Lagrangian (material), following
as it does the stretching history of fluid elements.  In the strongly chaotic
systems we deal with in the present context, Lagrangian and Eulerian
coordinates are related by a convoluted transformation.  This transformation
is so complex that its specific form is inaccessible (even numerically) after
some time, a reflection of sensitivity to initial conditions.  For long times,
the two frames must be regarded as essentially independent: we cannot simply
solve a problem in Eulerian coordinates and transform to Lagrangian
coordinates (or vice-versa).  Thus we believe it is worthwhile to take a
chaotic system whose solution has already been obtained in Eulerian
coordinates and solve it in Lagrangian coordinates.  As indicated above, this
links two views of mixing together, and in particular illustrates the nature
of strange eigenmodes in Lagrangian coordinates.  This also indicates the
source of the breakdown of local theories.  We will show that there exists a
kind of \emph{Lagrangian strange eigenmode}, which is not quite an eigenmode
but which exhibits similar features: specifically, it is an eigenmode if an
appropriate time-dependent rescaling of coordinates is performed (exponential
in time).  This rescaling is closely related to the ``cone'' involved in
diffusive problems in the presence of an exponentially stretching
flow~\cite{Zeldovich1984}.  We call it the \emph{cone of safety}, because
modes inside it are sheltered from diffusion at a given time.

We introduce the system to be studied, the perturbed cat map, in
Section~\ref{sec:torusmap}.  We find its finite-time Lyapunov exponents and
eigenvectors using first-order perturbation theory.  In Section~\ref{sec:AD}
we partially solve the \adeq\ for our map, again using perturbation theory.
Some numerical work is needed to complete the solution, and this is described
in Section~\ref{sec:numres}.  Finally, a few concluding remarks are offered in
Section~\ref{sec:disc}.

\section{The Perturbed Cat Map}
\label{sec:torusmap}

\begin{figure}
\psfrag{x1}{$\xc_1$}
\psfrag{x2}{\raisebox{.5em}{$\xc_2$}}
\psfrag{(a)}{(a)}
\psfrag{(b)}{(b)}
\psfrag{(c)}{(c)}
\psfrag{(d)}{(d)}
\includegraphics[width=.8\columnwidth]{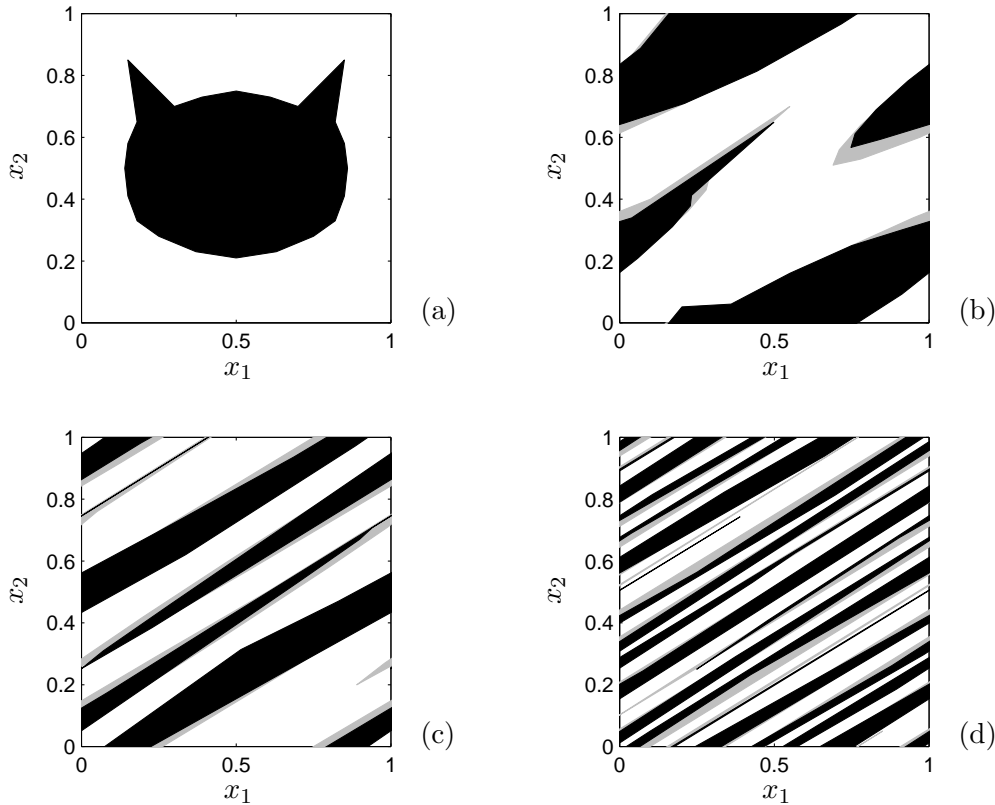}
\caption{Action of the perturbed cat map for~$\eps=0.4$. (a) Initial pattern;
  (b) First iterate; (c) Second iterate; (d) Third iterate.  The gray shading
  shows the action of the unperturbed cat map.}
\label{fig:catmap}
\end{figure}

The strange eigenmode has for the most part been studied in maps, because
these present great advantages for analytical work.  This is also reasonable
since experimental work has so far focused on time-periodic
flows~\cite{Rothstein1999,Voth2003}.  As in~\cite{Thiffeault2003d}, we
consider the map
\begin{equation}
  \M(\xv) = \ML\cdot\xv + \bm{\phi}(\xv),
  \label{eq:torusdiffeo}
\end{equation}
defined on the unit two-torus,~\hbox{$\twotorus = [0,1]^2$}.  Here~$\ML$ is a
matrix of integers with unit determinant, and~$\bm{\phi}(\xv)$ is a
doubly-periodic function, so that~$\M$ is a diffeomorphism; specifically, we
take
\begin{equation}
  \ML = \begin{pmatrix}2 & 1 \\ 1 & 1 \end{pmatrix};\qquad
  \bm{\phi}(\xv) = \frac{\eps}{2\pi}
  \begin{pmatrix}\sin2\pi\xc_1 \\ \sin2\pi\xc_1 \end{pmatrix};
  \label{eq:Mphi}
\end{equation}
where~$\bm{\phi}(\xv)$ is chosen such that~$\M$ is area-preserving.
For~$\eps=0$, \eqref{eq:torusdiffeo} is the usual cat map of
Arnold~\cite{Arnold}.  The map~\eqref{eq:torusdiffeo} inherits much of the
simplicity of the cat map, but the perturbation allows for more complex---and
less singular---behavior.  The action of the map is depicted in
Fig.~\ref{fig:catmap}.  For small~$\eps$, the map is very close to the cat
map, but the implications of the perturbation for mixing are profound, as we
will now discuss.

We are interested in the mixing properties of the map~$\M$.  In
Ref.~\cite{Thiffeault2003d} mixing in this map was investigated from an
Eulerian perspective: advection alternated with diffusion and the central
object was the distribution of the concentration field in Eulerian
coordinates.  For~$\eps=0$, mixing occurs superexponentially in the map, due
to the lack of dispersion in Fourier space.  The concentration in a given
Fourier mode is mapped entirely to one mode of higher wavenumber, and so on to
ever higher wavenumbers.  This sequence of wavenumbers have
exponentially-growing magnitudes.  Because diffusion is exponential in the
wavenumber, the net result is superexponential decay (\ie, the exponential of
minus an exponential in time).

For~$\eps\ne 0$, the situation is radically different.  The map now disperses
concentration among many Fourier modes at each iteration.  In particular, some
concentration is always mapped back to the lowest allowable wavenumber (the
grave mode), on which the weak diffusion is ineffective.  The decay of the
scalar is then limited by how much concentration is mapped to this grave mode
at each iteration.  The grave mode thus forms the seed of a strange eigenmode,
since an eigenmode is by definition a \emph{recurring} feature.  The
concentration field thus settles into the slowest-decaying eigenfunction of
the advection--diffusion operator---the strange eigenmode---analogously to
earlier
work~\cite{Pierrehumbert1994,Fereday2002,Wonhas2002,Sukhatme2002,Pikovsky2003,%
Liu2004}.

Here we wish to solve the same problem as in Ref.~\cite{Thiffeault2003d}, but
in Lagrangian coordinates.  This means that we focus on following fluid
elements and describing how they deform under the action of the map.  To solve
the advection--diffusion problem in Lagrangian coordinates, it is necessary to
have expressions for the finite-time Lyapunov exponents of the
map~\cite{Thiffeault2003b} (or equivalently the coefficients of expansion)
and their associated characteristic directions, as a function of Lagrangian
coordinates, and not just their distribution (we will see why this is so in
Section~\ref{sec:AD}).  Because the finite-time Lyapunov exponents are easily
derived for the cat map ($\eps=0$), we shall proceed perturbatively,
assuming~$\eps$ is small.

First let us give the Lyapunov exponents and associated characteristic
directions for the unperturbed cat map.  The Lyapunov exponents are the
logarithms of the eigenvalues of~$\ML$, and the characteristic directions are
the corresponding eigenvectors.  It is convenient to introduce an
angle~$\theta$ in terms of which the eigenvectors of~$\ML$ are
\begin{equation}
  (\udiruv\ \ \sdiruv) = \begin{pmatrix}
    \cos\theta & -\sin\theta\\
    \sin\theta & \cos\theta
  \end{pmatrix}
  \label{eq:edirv}
\end{equation}
with~$\cos^2\theta = \tfrac{1}{2}(1 + 1/\sqrt{5})$ and~$\sin^2\theta =
\tfrac{1}{2}(1 - 1/\sqrt{5})$.  Then the corresponding eigenvalues of~$\ML$
are
\begin{subequations}
\begin{align}
  \nugr_\udir = \nugr &= \tfrac{1}{2}(3 + \sqrt{5}) = 1 + \cot\theta
  = \cot^2\theta,
  \label{eq:nugru}\\
  \nugr_\sdir = \nugr^{-1} &= \tfrac{1}{2}(3 - \sqrt{5}) = 1 - \tan\theta
  = \tan^2\theta.
  \label{eq:nugrs}
\end{align}
\label{eq:nugr}
\end{subequations}
These equalities are specific to this particular angle, as is the
relation~$\tan\theta = \cot\theta-1$.  The~$\udiruv$ direction is associated
with stretching, and~$\sdiruv$ with contraction.

The coefficients of expansion (given by~$\nugr^\iter$ and~$\nugr^{-\iter}$
after~$\iter$ iterations of the map) and characteristic directions for the
linear cat map are uniform in space.  Now we derive their value for~$\eps$
nonzero but small, using perturbation theory.  The problem is to find the
eigenvalues and eigenvectors of the matrix~$\metric^{(\iter)}$, with
components
\begin{equation}
  \metric^{(\iter)}_{pq} \ldef \sum_\iterj
  \frac{\pd\xc^{(\iter)}_\iterj}{\pd\Xc_p}\,
		\frac{\pd\xc^{(\iter)}_\iterj}{\pd\Xc_q}\,,
	\label{eq:metric}
\end{equation}
often called the metric tensor (or Cauchy--Green strain tensor in fluid
mechanics).  Here~$\Xv$ is the Lagrangian label (coordinate) and~$\xv =
\xv^{(\iter)}(\Xv)$ is the~$\iter$th iterate of the point~$\Xv$ under the
action of the map~\eqref{eq:torusdiffeo} (so
that~\hbox{$\xv^{(0)}(\Xv)=\Xv$}).  The metric tensor describes the stretching
experienced at the~$\iter$th iteration by a fluid element initially at~$\Xv$.
Its eigenvalues and eigenvectors give the shape and orientation at
the~$\iter$th iteration of an ellipsoid representing an initially spherical
infinitesimal element of fluid.

Before we can apply perturbation theory to the metric tensor, we must find the
form of the perturbation itself.  To first order in~$\eps$, the~$\iter$th
iterate of the map~\eqref{eq:torusdiffeo} is
\begin{equation}
  \xv^{(\iter)} = \M^\iter(\Xv) = \ML^\iter\cdot\Xv
  + \sum_{\iterj=0}^{\iter-1}\ML^\iterj\cdot\bm{\phi}(\ML^{\iter-\iterj-1}\Xv).
  \label{eq:xpert}
\end{equation}
The Jacobian matrix of this transformation is
\begin{equation}
  \frac{\pd\xv^{(\iter)}}{\pd\Xv} = \ML^\iter
  + \sum_{\iterj=0}^{\iter-1}\ML^\iterj\cdot
  \frac{\pd\bm{\phi}}{\pd\xv}(\ML^{\iter-\iterj-1}\Xv)
  \cdot\ML^{\iter-\iterj-1}
  \label{eq:dxpert}
\end{equation}
where the numerator corresponds to rows and the denominator to columns of a
matrix.  We must now construct the metric tensor~\eqref{eq:metric}, which to
leading order in~$\eps$ is
\begin{equation}
  \metric_\eps^{(\iter)} = \widetilde{\ML}^\iter\,\ML^\iter
  + \bigl\{\gpert^{(\iter)} + \widetilde{\gpert^{(\iter)}}\bigr\}
  \label{eq:gpert}
\end{equation}
with
\begin{equation}
  \gpert^{(\iter)} \ldef
  \widetilde{\ML}^\iter\cdot\sum_{\iterj=0}^{\iter-1}\ML^\iterj\cdot
  \frac{\pd\bm{\phi}}{\pd\xv}(\ML^{\iter-\iterj-1}\Xv)
  \cdot\ML^{\iter-\iterj-1}
  \label{eq:hpert}
\end{equation}
and the tilde denotes the transpose of a matrix.  The unperturbed metric
is~\hbox{$\metric_0^{(\iter)} = \widetilde{\ML}^\iter\,\ML^\iter$}, with
eigenvalues~$\nugr^{2\iter}$ and~$\nugr^{-2\iter}$ and eigenvectors given
by~\eqref{eq:edirv}.  The perturbation is the bracketed term
in~\eqref{eq:gpert}.  Finding the eigenvalues and eigenvectors of the
symmetric matrix~$\metric^{(\iter)}$ to leading order in~$\eps$ is a
straightforward application of perturbation theory for symmetric matrices,
familiar from quantum mechanics.  For more details we refer the reader to
standard texts on the subject~\cite{Kato,Merzbacher}.

To leading order in~$\eps$, the coefficient of stretching is written as
\begin{equation}
  \nugr^{(\iter)}_\eps(\Xv) = \nugr^\iter\,(1 + \eps\eta^{(\iter)}(\Xv))
\end{equation}
where~$\nugr^\iter$ is the coefficient of stretching of the unperturbed cat
map, and the correction~\hbox{$\nugr^\iter\eps\eta^{(\iter)}(\Xv)$} is
obtained from the perturbation by contraction with the unperturbed
eigenvectors~\cite{Kato,Merzbacher},
\begin{equation}
  \nugr^\iter\eps\eta^{(\iter)}(\Xv) = \tfrac{1}{2}\,
  \udiruv\cdot\bigl\{\gpert^{(\iter)} + \widetilde{\gpert^{(\iter)}}
  \bigr\}\cdot\udiruv
  = \udiruv\cdot\gpert^{(\iter)}\cdot\udiruv\,.
  \label{eq:scpert}
\end{equation}
Observe that because the coefficient of stretching is the square root of the
largest eigenvalue of~$\metric_\eps^{(\iter)}$, there is an extra factor
of~$1/2$ to leading order in~$\eps$.  Using the fact that, for~$\ML$
symmetric,~$\ML\cdot\udiruv=\udiruv\cdot\ML=\nugr\,\udiruv$,
we find Eqs.~\eqref{eq:hpert} and~\eqref{eq:scpert} give
\begin{equation}
  \eta^{(\iter)} = \sin\theta\cos\theta\sum_{\iterj=0}^{\iter-1}
    \cos\l(2\pi(\ML^\iterj\cdot\Xv)_1\r)
  \label{eq:etadef}
\end{equation}
where we have substituted the specific form of the map, given
by~\eqref{eq:Mphi}.  The subscript `1' in~\eqref{eq:etadef} indicates the
$\xc_1$ component of a vector.

The perturbed eigenvectors can be written as
\begin{equation}
  \udiruv_\eps^{(\iter)}(\Xv) = \udiruv
  + \eps\zeta^{(\iter)}(\Xv)\,\sdiruv\,,
  \qquad
  \sdiruv_\eps^{(\iter)}(\Xv) = \sdiruv
  - \eps\zeta^{(\iter)}(\Xv)\,\udiruv\,,
  \label{eq:edirpert}
\end{equation}
where~$\eps\zeta^{(\iter)}$ may be regarded as a small angle of rotation.
Again we follow standard matrix perturbation theory~\cite{Kato,Merzbacher}, so
that the angle of rotation is given by
\begin{equation}
  \eps\zeta^{(\iter)}(\Xv) = \frac{\udiruv\cdot\bigl\{\gpert^{(\iter)} +
  \widetilde{\gpert^{(\iter)}}\bigr\}\cdot\sdiruv}{\nugr^{2\iter}
    - \nugr^{-2\iter}}\,,
\end{equation}
which after some reduction and the use of~\eqref{eq:Mphi} yields
\begin{equation}
  \zeta^{(\iter)} = \frac{1}{\nugr^{2\iter} - \nugr^{-2\iter}}\,
  (\zeta_+^{(\iter)}  + \zeta_-^{(\iter)}),
  \label{eq:zetadef}
\end{equation}
with
\begin{equation}
  \zeta_\pm^{(\iter)} \ldef
  \tfrac{1}{2}(\cos2\theta\mp1)
  \sum_{\iterj=0}^{\iter-1}\nugr^{\pm2(\iter-\iterj)}
  \cos\l(2\pi(\ML^\iterj\cdot\Xv)_1\r).
  \label{eq:zetapmdef}
\end{equation}
Note that the asymptotic direction ($\iter\gg 1$) is dominated
by~$\zeta_+^{(\iter)}$, so that
\begin{equation}
  \zeta^{(\iter)} \simeq
  \cos^2\theta
  \sum_{\iterj=0}^{\iter-1}\nugr^{-2\iterj}
  \cos\l(2\pi(\ML^\iterj\cdot\Xv)_1\r),
  \qquad \iter\gg 1.
  \label{eq:zetainf}
\end{equation}
In this form it is easy to check that~$\udiruv\cdot\grad\zeta^{(\iter)} =
\sdiruv\cdot\grad\eta^{(\iter)}$ for $\iter\gg 1$, as required by the
differential constraint \hbox{$\div\sdiruv_\eps^{(\iter)} +
\sdiruv_\eps^{(\iter)}\cdot\grad\log\nugr_\eps^{(\iter)} =
0$}~\cite{Tang1996,Thiffeault2001,Thiffeault2002}. (The derivatives are taken
with respect to the Lagrangian coordinates~$\Xv$.)

\begin{figure}
\psfrag{eta}{$\eta^{(\iter)}$}
\psfrag{iter}{$\iter$}
\psfrag{eps = 1.0e-05}{$\eps = 10^{-5}$}
\includegraphics[width=.8\columnwidth]{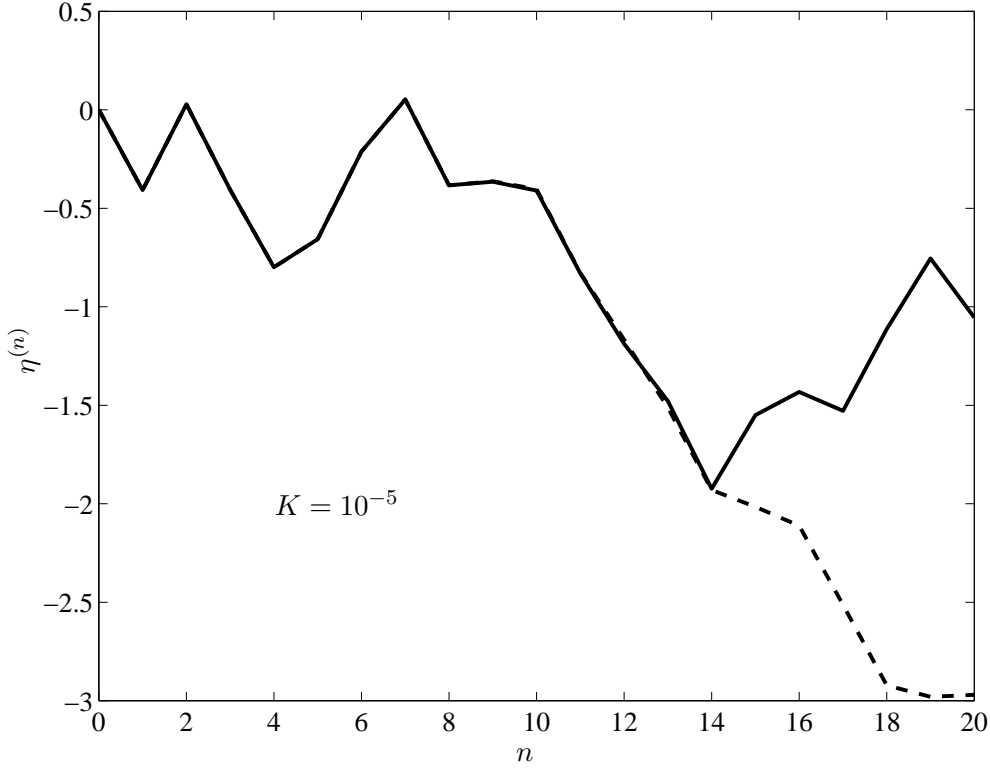}
\caption{Numerical solution (solid) and first-order solution~$\eta^{(\iter)}$
(dashed) from Eq.~\eqref{eq:etadef} for~\hbox{$\eps=10^{-5}$}.  The solutions
diverge after several iterations because we are perturbing off a chaotic
trajectory.}
\label{fig:torusmap_lyap}
\end{figure}

\begin{figure}
\psfrag{sx}{\hspace{-1em}\raisebox{1em}
  {$\Bigl(\sdiruv^{(\infty)}_\eps\Bigr)_1$}}
\psfrag{x}{\raisebox{-.4em}{$\Xc_1$}}
\psfrag{eps = 1.0e-01}{$\eps = 0.1$}
\includegraphics[width=.8\columnwidth]{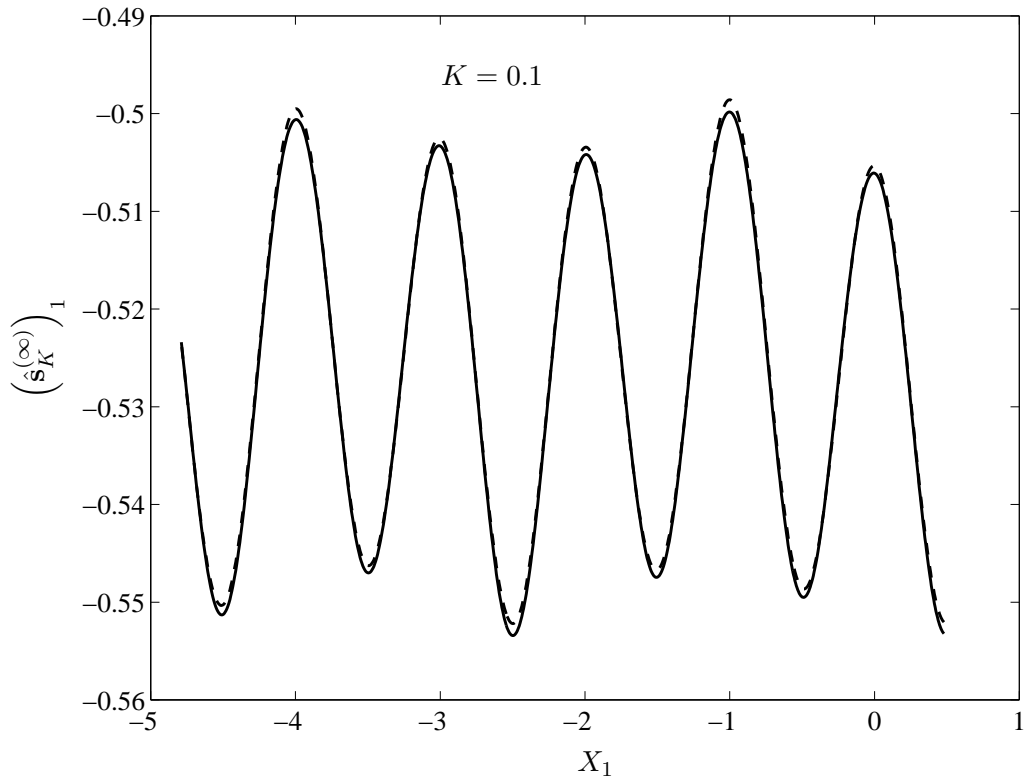}
\caption{Numerical solution (solid) and first-order solution (dashed) for the
first component of the asymptotic eigenvector~$\sdiruv^{(\infty)}_\eps$
with~$\eps=0.1$, using the asymptotic result~\eqref{eq:zetainf}
in~\eqref{eq:edirpert}.  The error is of order~$\eps^2$.}
\label{fig:torusmap_s}
\end{figure}

The first-order perturbative solution for the coefficient of
stretching~$\eta^{(\iter)}$ is compared to numerical results in
Fig.~\ref{fig:torusmap_lyap}, and similarly for the
eigenvector~$\sdiruv^{(\iter)}_\eps$ in Fig.~\ref{fig:torusmap_s}.
Unlike the perturbed coefficients of stretching, which eventually diverge from
the numerical solution because of the sensitivity to initial conditions, the
perturbed eigenvectors converge very rapidly and are always close to the
numerical result.

It will be more convenient to express the metric tensor in terms of its
eigenvalues and eigenvectors.  The metric tensor~\eqref{eq:metric} can be
written in terms of the coefficients of expansion and characteristic
directions as
\begin{equation}
  \metric^{(\iter)}
  = [\nugr^{(\iter)}]^2\udiruv^{(\iter)}\udiruv^{(\iter)}
  + [\nugr^{(\iter)}]^{-2}\sdiruv^{(\iter)}\sdiruv^{(\iter)}\,.
\end{equation}
To leading order in~$\eps$, we have
\begin{equation}
  \metric_\eps^{(\iter)} = \nugr^{2\iter}\,\udiruv\,\udiruv
  + \nugr^{-2\iter}\,\sdiruv\,\sdiruv
  + 2\eps\,\eta^{(\iter)}(\nugr^{2\iter}\,\udiruv\,\udiruv -
  \nugr^{-2\iter}\,\sdiruv\,\sdiruv)
  + \eps\,\zeta^{(\iter)}\l(\nugr^{2\iter} - \nugr^{-2\iter}\r)
  (\udiruv\,\sdiruv + \sdiruv\,\udiruv).
  \label{eq:geps}
\end{equation}
The only dependence on~$\Xv$ in~\eqref{eq:geps} is contained
in~$\eta^{(\iter)}$ and~$\zeta^{(\iter)}$.

\section{Advection and Diffusion}
\label{sec:AD}

Having derived the coefficients of expansion and characteristic directions of
stretching (to leading order in~$\eps$), we can now solve the \adeq\ in
Lagrangian coordinates.  We will first discuss the case for an incompressible
flow, and then make the transition to a volume-preserving map.

The \adeq\ for the concentration of a
scalar,~$\Theta(\xv,\time)$, advected by an incompressible velocity
field~$\velv$ is
\begin{equation}
	\pd_\time\Theta
		+ \velv\cdot\pd_\xv\Theta = \Diffc\,\pd_\xv^2\Theta,
	\label{eq:adeq}
\end{equation}
where~$\Diffc$ is the diffusion coefficient.  We define the
transformation~$\xv(\Xv,\time)$ from Lagrangian coordinates~$\Xv$ to Eulerian
coordinates~$\xv$ by
\begin{equation}
  \dot\xv = \velv(\xv,\time),\qquad \xv(\Xv,0)=\Xv,
  \label{eq:L2E}
\end{equation}
where the overdot denotes a time derivative at fixed~$\Xv$.  We can then
transform~\eqref{eq:adeq} to Lagrangian
coordinates~$\Xv$~\cite{Tang1996,Thiffeault2003b},
\begin{equation}
	\dot\Theta = \pd_\Xv\cdot(\Difft\cdot\pd_\Xv\Theta),
	\label{eq:adeqLagr}
\end{equation}
where we reused the same symbol for~$\Theta(\Xv,\time)$.
The anisotropic, nonhomogeneous, time-dependent diffusion tensor~$\Difft$ is
\begin{equation}
	\Difft \ldef \Diffc\,\metric^{-1},
	\qquad
	\metric_{pq} \ldef \sum_\iterj \frac{\pd\xc_\iterj}{\pd\Xc_p}\,
		\frac{\pd\xc_\iterj}{\pd\Xc_q}\,,
	\label{eq:Difft}
\end{equation}
where~$\metric$ is the metric tensor encountered in
Section~\ref{sec:torusmap}.  By construction, the advection term has
disappeared from~\eqref{eq:adeqLagr}.  The flow enters~\eqref{eq:adeqLagr}
indirectly through the metric tensor in~\eqref{eq:Difft}, reflecting the
enhancement to diffusion due to the deformation of fluid
elements~\cite{Tang1996,Thiffeault2003b}.

We now make the leap from a flow to a map: because the velocity field does not
enter~\eqref{eq:adeqLagr} directly, we may regard the time dependence
in~$\Difft$ as given by a map rather than a flow, and use the
metric~\eqref{eq:metric} in the diffusion tensor~$\Difft$~\footnote{It is not
possible to simply invoke an incompressible flow corresponding to a map: in
general there is no incompressible flow of the same dimension whose
trajectories agree with a given map.  Formally, however, the replacement of
the metric tensor by one corresponding to a map is well-defined
mathematically.}.  We also write~$\Theta^{(\iter)}(\Xv)$
for~$\Theta(\Xv,\time)$, where~$\iter$ denotes the~$\iter$th iterate of the
map.

Since our map is defined on the torus, we can expand~$\Theta^{(\iter)}(\Xv)$
in Fourier components~$\widehat\Theta^{(\iter)}_{\kv}$; the resulting map,
obtained by first Fourier transforming and then solving~\eqref{eq:adeqLagr},
is
\begin{equation}
  \widehat\Theta_{\kv}^{(\iter)} =
  \sum_{\lv}
  \exp\bigl(\G^{(\iter)}\bigr)_{\kv\lv}\widehat\Theta_{\lv}^{(\iter-1)}\,,
  \label{eq:ADmap}
\end{equation}
where
\begin{equation}
  \G^{(\iter)}_{\kv\lv}
  = -4\pi^2 \Time\int_\twotorus (\kv\cdot\Difft^{(\iter)}\cdot\lv)\,
  \ee^{-2\pi\imi(\kv-\lv)\cdot\xv} \dint^2 x\,,
  \label{eq:GFour}
\end{equation}
with~$\Time$ the period of the map.  This is an exact result, but the great
difficulty lies in calculating the exponential of~$\G^{(\iter)}$.  Again, we
shall accomplish this perturbatively.

For the torus map introduced in Section~\ref{sec:torusmap}, from
Eq.~\eqref{eq:geps} we obtain
\begin{equation}
  [\metric_\eps^{(\iter)}]^{-1} = \nugr^{2\iter}\,\sdiruv\,\sdiruv
  + \nugr^{-2\iter}\,\udiruv\,\udiruv
  + 2\eps\,\eta^{(\iter)}(\nugr^{2\iter}\,\sdiruv\,\sdiruv -
  \nugr^{-2\iter}\,\udiruv\,\udiruv)
  - \eps\,\zeta^{(\iter)}\l(\nugr^{2\iter} - \nugr^{-2\iter}\r)
  (\udiruv\,\sdiruv + \sdiruv\,\udiruv),
  \label{eq:ginveps}
\end{equation}
to leading order in~$\eps$, where the only functions of~$\Xv$
are~$\eta^{(\iter)}$ and~$\zeta^{(\iter)}$.  Inserting~\eqref{eq:ginveps}
into~\eqref{eq:GFour}, we find
\begin{equation}
  \G^{(\iter)}_{\kv\lv} = \A^{(\iter)}_{\kv\lv} + \eps \B^{(\iter)}_{\kv\lv}
  \label{eq:GAB}
\end{equation}
where
\begin{equation}
  \A^{(\iter)}_{\kv\lv} \ldef -\scaledDiffc\l(\nugr^{2\iter}\,k_\sdir^2
  + \nugr^{-2\iter}\,k_\udir^2\r)\delta_{\kv\lv}
  \label{eq:Adef}
\end{equation}
and
\begin{equation}
  \B^{(\iter)}_{\kv\lv} \ldef -\scaledDiffc
  \l(2\!\l(\nugr^{2\iter}\,k_\sdir\,\ell_\sdir
  - \nugr^{-2\iter}\,k_\udir\,\ell_\udir\r)\eta^{(\iter)}_{\kv\lv}
  - \l(k_\udir\,\ell_\sdir + k_\sdir\,\ell_\udir\r)
  (\zeta_+^{(\iter)}{}_{\kv\lv} + \zeta_-^{(\iter)}{}_{\kv\lv})\r).
  \label{eq:Bdef0}
\end{equation}
Here we have defined
\begin{equation}
  \scaledDiffc \ldef 4\pi^2 \Diffc\,\Time
\end{equation}
to agree with the notation in Ref.~\cite{Thiffeault2003d}, as well as
\begin{equation}
  k_\udir \ldef \kv\cdot\udiruv, \qquad k_\sdir\ldef\kv\cdot\sdiruv,
\end{equation}
and similarly for~$\ell_\udir$ and~$\ell_\sdir$.

Upon making use of the
Fourier-transformed~\eqref{eq:etadef},~\eqref{eq:zetadef},
and~\eqref{eq:zetapmdef} in~\eqref{eq:Bdef0}, we find
\begin{equation}
  \B^{(\iter)}_{\kv\lv} = -\tfrac{1}{2}\scaledDiffc
  \sum_{\iterj=0}^{\iter-1} \mathcal{B}^{\iter \iterj}_{\kv\lv}
  \l(\delta_{\kv , \lv + \uvx\cdot\ML^\iterj}
  + \delta_{\kv , \lv - \uvx\cdot\ML^\iterj}\r)
  \label{eq:Bdef}
\end{equation}
where~$\uvx$ is a unit vector in the~$\xc_1$ direction, and
\begin{equation}
  \mathcal{B}^{\iter \iterj}_{\kv\lv} \ldef \sin2\theta
  \l(\nugr^{2\iter}\,k_\sdir\,\ell_\sdir
  - \nugr^{-2\iter}\,k_\udir\,\ell_\udir\r)
  + \l(k_\udir\,\ell_\sdir + k_\sdir\,\ell_\udir\r)
  \l(\nugr^{2(\iter-\iterj)}\sin^2\theta
  - \nugr^{-2(\iter-\iterj)}\cos^2\theta\r).
\end{equation}

To obtain the full solution, we must now exponentiate~\eqref{eq:GAB} to give
the transfer matrix in~\eqref{eq:ADmap}.  Fortunately, for~$\A$ diagonal
there is a simple expansion,
\begin{equation}
  [\exp(\A^{(\iter)} + \eps \B^{(\iter)})]_{\kv\lv} =
  \ee^{\A^{(\iter)}_{\kv\kv}}\,\delta_{\kv\lv} + \eps \E^{(\iter)}_{\kv\lv};
  \qquad
  \E^{(\iter)}_{\kv\lv} \ldef
  \B^{(\iter)}_{\kv\lv}\,\frac{\ee^{\A^{(\iter)}_{\kv\kv}} -
  \ee^{\A^{(\iter)}_{\lv\lv}}}
  {\A^{(\iter)}_{\kv\kv} - \A^{(\iter)}_{\lv\lv}}\,,
  \label{eq:Edef}
\end{equation}
valid to first order in~$\eps$.  We say fortunately because without such a
formula it is very difficult to compute this matrix exponential---even
numerically---due to the large size of the matrices (\ie, infinite) and their
magnitude (\ie, growing exponentially in time).

The~$\nugr^{2\iter}$ term in~$\A_{\kv\kv}^{(\iter)}$ seems to imply
that~$\widehat\Theta^{(\iter)}_{\kv}$ decays \emph{superexponentially} fast as
$\exp(-\scaledDiffc\,\nugr^{2\iter}\,k_\sdir^2)$.  From Eulerian
considerations~\cite{Thiffeault2003d}, we know that for~\hbox{$\eps\ne 0$} the
decay is actually exponential after a short superexponential transient.  This
is because the~$\E^{(\iter)}$ term must be taken into account: it breaks the
diagonality of~$\G^{(\iter)}$, so that given some initial set of wavevectors,
the concentration contained in those modes can be transferred elsewhere.
In particular, it can transfer concentration to modes aligned with the
unstable direction.  We will see how this avoids superexponential decay in
Section~\ref{sec:numres}.

\section{Numerical Results}
\label{sec:numres}

\subsection{The Numerical Method}
\label{sec:nummet}

At this point, solving~\eqref{eq:ADmap} and~\eqref{eq:Edef} numerically seems
like the only way forward.  Clearly, attempting the solve this on a grid in
Fourier space is hopeless: very high wavenumber modes are quickly populated so
the resolution is exhausted very rapidly.  Instead, the procedure we use
involves keeping track of a list of excited Fourier modes (\ie, those that are
nonzero to machine precision).  We now describe this scheme.

First, an initial wavenumber is seeded with some initial concentration.  This
mode will be damped by the diagonal part of the matrix in~\eqref{eq:Edef}, and
will also excite two new modes as seen in~\eqref{eq:Bdef}.  Repeating this,
starting now from three modes, we see that the number of excited modes grows
exponentially.  Thus it would seem that this procedure is not very
advantageous; however, after a few iteration the diffusivity (the diagonal
part in~\eqref{eq:Edef}) will damp most modes because~$\A^{(\iter)}_{\kv\kv}$
is growing exponentially.  Thus the modes that have been damped beyond
redemption can be removed from the list.  In this manner the number of excited
modes eventually reaches a constant, though they consist of ever higher
wavenumbers.  Thus one can think of a ``packet'' of modes cascading through
Fourier space towards larger wavenumbers.  It is this packet that is the
Lagrangian analogue of the strange eigenmode in Eulerian space, as we will
discuss in Section~\ref{sec:Lagrstr}.

Let us first present some results.
\begin{figure}
\psfrag{eps = 1.0e-03}{$\eps = 10^{-3}$}
\psfrag{iter}{\raisebox{-.25em}{$\iter$}}
\psfrag{variance}{\raisebox{.5em}{variance}}
\psfrag{D=0.0001}{$\scriptstyle\scaledDiffc\ =\ 10^{-4}$}
\psfrag{D=0.001 }{$\scriptstyle\scaledDiffc\ =\ 10^{-3}$}
\psfrag{D=0.01  }{$\scriptstyle\scaledDiffc\ =\ 10^{-2}$}
\psfrag{D=0.1   }{$\scriptstyle\scaledDiffc\ =\ 10^{-1}$}
\includegraphics[width=.8\columnwidth]{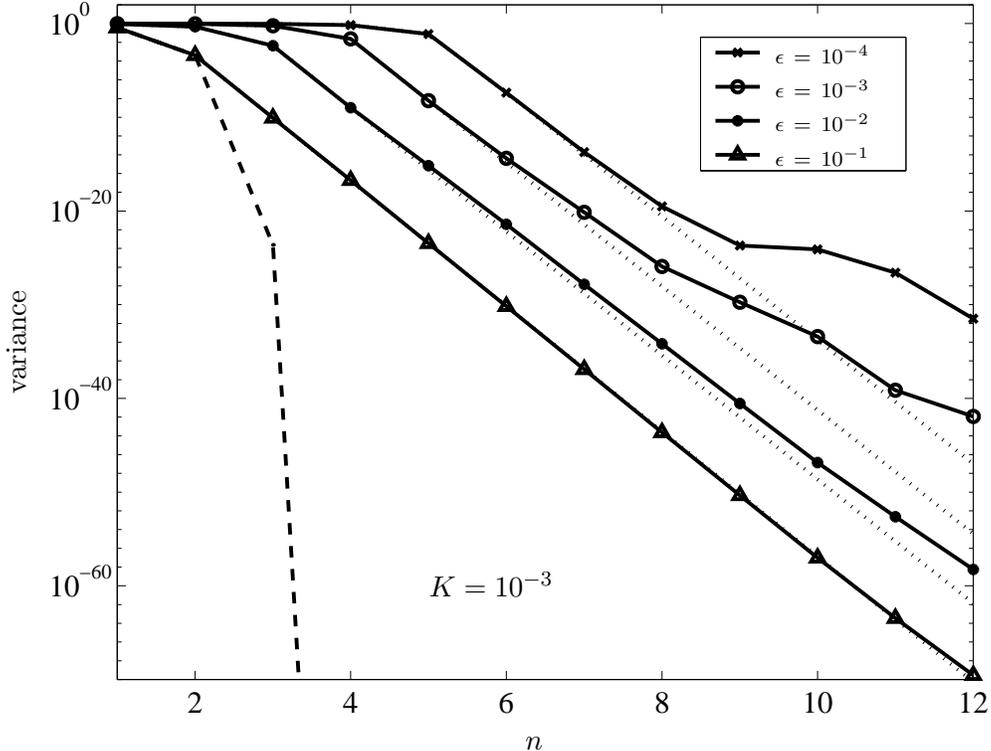}
\caption{Decay of the variance for~$\eps=0.001$ and different values
of~$\scaledDiffc$, compared to the result from Eulerian coordinates
(dotted lines).  The dashed line shows the exact result for superexponential
decay ($\eps=0$) for~$\scaledDiffc=0.1$.}
\label{fig:var_eps=0.001}
\end{figure}
Figure~\ref{fig:var_eps=0.001} shows the decay of the scalar variance
for~\hbox{$\eps=0.001$} and four values of the diffusivity.  The agreement
with the Eulerian results is excellent for early times, but inevitably breaks
down later. (In fact the variance eventually begins to \emph{increase}, which
is forbidden.)  The agreement is also worse for smaller diffusivity.  Both of
these disagreements are a manifestation of the wavenumber dependence of the
perturbation in~\eqref{eq:Edef}: for~$k$ too large the perturbation becomes
large, invalidating the approach.  Nevertheless, Fig.~\ref{fig:var_eps=0.001}
clearly validates the calculation for times that are not too long.

\begin{figure}
\psfrag{eps}{$\eps$}
\psfrag{delta}{\raisebox{.5em}{\hspace{-1em}Difference}}
\psfrag{D=0.01}{$\scaledDiffc = 10^{-2}$}
\psfrag{iter=4}{$\iter = 4$}
\psfrag{eps2}{$\eps^2$}
\includegraphics[width=.8\columnwidth]{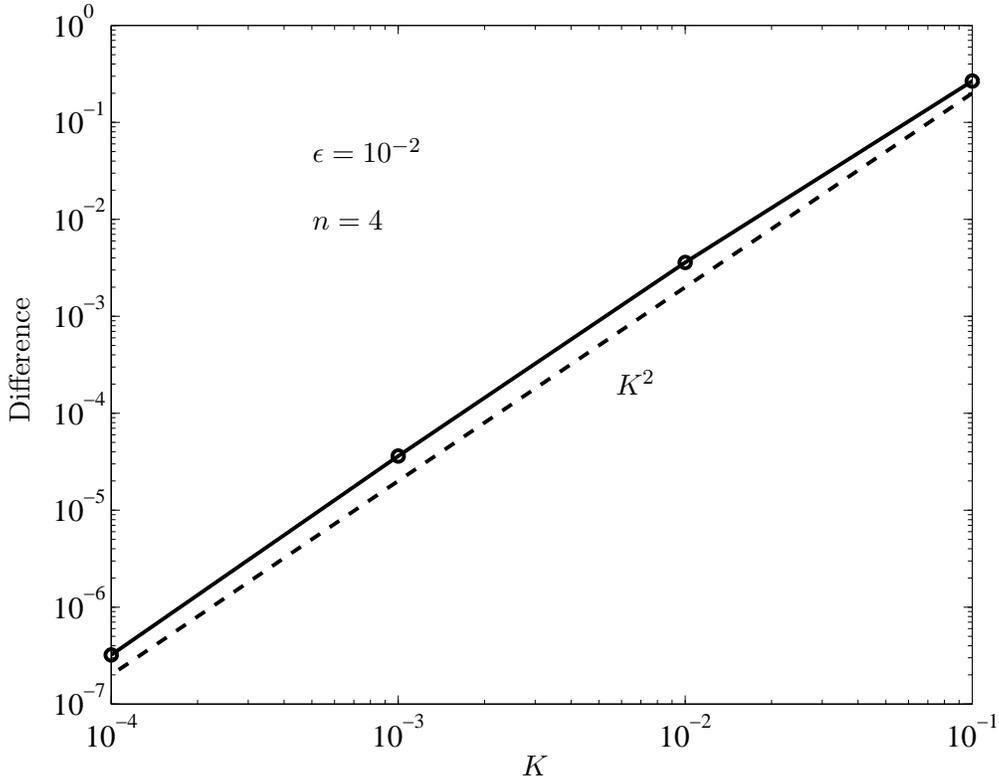}
\caption{Relative difference in the variance between the Eulerian and
perturbative Lagrangian result as a function of~$\eps$, at the fourth
iteration.  The dashed line indicates an~$\eps^2$ dependence, showing that the
two agree to first order in~$\eps$.}
\label{fig:converg}
\end{figure}
Another validation is shown in Fig.~\ref{fig:converg}, where we plot the
difference between the Eulerian and Lagrangian results as a function of
$\eps$.  The difference clearly scales as~$\eps^{-2}$, showing that the two
agree at leading order, as required for a first-order asymptotic result.

We now interpret our results in greater detail, and look for a manifestation
of the strange eigenmode in Lagrangian coordinates.

\subsection{The Lagrangian Strange Eigenmode}
\label{sec:Lagrstr}

The mechanism described in Section~\ref{sec:nummet} is similar to that
originally introduced (in the context of the kinematic dynamo problem) by
Zeldovich~\etal~\cite{Zeldovich1984}: they basically solved the \adeq\ in
Lagrangian coordinates for a linear velocity field, and found that in order to
avoid rapid superexponential decay one had to restrict attention to a ``cone''
of wavenumbers that are closely aligned with the unstable manifold of the flow
(a similar approach was used later in
Refs.~\cite{Antonsen1996,Son1999,Balkovsky1999}).  The exponential shrinking
in time of this ``cone of safety'' leads to an exponential decay of scalar
variance at a rate given by the Lyapunov exponents.

The problem with that approach is that a linear velocity field offers no
possibility of \emph{dispersion} in Fourier space.  The wavenumbers in the
cone of safety must have some concentration associated with them initially.
What our numerical results show is that if one considers dispersion in Fourier
space (of the type allowed by the~$\E^{(\iter)}$ term in~\eqref{eq:Edef}) then
it is possible for concentration to be moved inside the cone from elsewhere.
The Lagrangian equivalent of the strange eigenmode must live within the cone
of safety, otherwise it would decay away superexponentially.  But unlike
Ref.~\cite{Zeldovich1984} the decay rate in the present case is not determined
by the shrinking of the cone: it is set by how much variance gets transferred
into the cone at each iteration.

\begin{figure}
\psfrag{log10\(scaled amplitude\)}{\raisebox{.5em}{$\log_{10}$ of
    scaled amplitude}}
\psfrag{k \(scaled by L\(i-2\)*sint\)}
       {\raisebox{-.5em}{\hspace{3.6em}$\lVert\kv\rVert/k_\udir^{(\iter-2)}$}}
\psfrag{eps=1.0e-4}{$\!\!\!\eps = 10^{-4}$}
\psfrag{D=1.0e-1}{$\scaledDiffc = 10^{-1}$}
\includegraphics[width=.8\columnwidth]{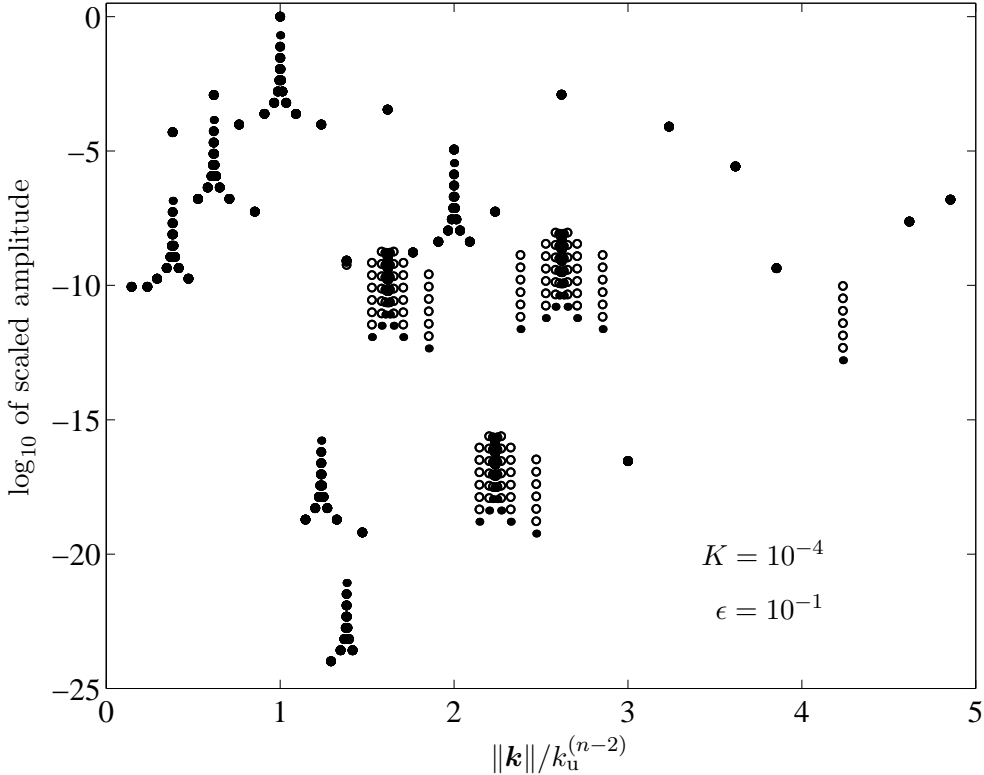}
\caption{The power spectrum of the Lagrangian strange eigenmode
  for~$\iter=6,\ldots,11$ (circles) and~$\iter=12$ (black dots).  The large
  black dots are points that are the same for all iterations (after
  rescaling): this is the dominant strange eigenmode in Lagrangian
  coordinates.  Both axes have been rescaled such that the dominant peak has
  unit amplitude and wavenumber ($k_\udir^{(\iter-2)}$ is defined in
  Eq.~\eqref{eq:kaction}).  The hollow circles are due to an admixture of
  another, faster-decaying eigenfunction.}
\label{fig:spectrum}
\end{figure}

Figure~\ref{fig:spectrum} shows a plot of the power spectrum of concentration.
The magnitude of the concentration ($\log_{10}$) is plotted vs the magnitude
of the wavenumber normalized by~$\lVert\kv\rVert_{\mathrm{max}}$ (its maximum
value), which is proportional to~$\nugr^\iter$.  The concentration is
normalized at each iteration such that the mode with largest concentration has
unit magnitude.  The iterations plotted are~$\iter=6,7,8,9,10,11$ (circles)
and~$\iter=12$ (black dots).  Most of the circles appear as large black dots,
because all these points lie on top of each other.  Hence, the concentration
is in an eigenstate, given that the wavenumber has been rescaled by a factor
proportional to~$\nugr^{\iter}$ (\ie, such that the dominant peak is at unit
rescaled wavenumber).  This is what we interpret as the Lagrangian equivalent
of the strange eigenmode (a good name might be ``stretched eigenmode'' in
Fourier space).

Some points in Fig.~\ref{fig:spectrum} exhibit a decay with iteration number
(appearing as columns of circles, with higher iteration numbers lower on the
graph): they belong to a more rapidly-decaying eigenfunction.  Note that the
peaks do not sharpen with iteration number, but more points are added to some
of the tails.  The eigenfunction appears extremely rough and discontinuous,
though the peaks are indicative of some underlying continuum behavior.  The
seemingly isolated points actually tend to line up with a peak far below.
Finally, note that the relative height (but not position or shape) of the
peaks depends on~$\eps$: the whole shape is stretched vertically as~$\eps$ is
made smaller.  This is because the term proportional to~$\eps$ controls the
transfer of concentration ``vertically'' (with respect to
Fig.~\ref{fig:spectrum}) in the eigenmode at each iteration.

\section{Discussion}
\label{sec:disc}

The Lagrangian strange eigenmode has some intriguing features: (i) It is
rescaled exponentially in time, in order to remain within the cone of safety
(so it is not a true eigenmode); (ii) Its power spectrum is very
discontinuous, in sharp contrast to its Eulerian
counterpart~\cite{Thiffeault2003d}; (iii) Its decay rate is set by how much
concentration is moved into the ``new'' cone of safety at each iteration
(since the cone is shrinking).  In the Appendix we present an analytic result
for a two-mode system which gives a simplified representation of the cone of
safety.

In the map analyzed here the exponential time-rescaling gives a proper
eigenmode, since only the constant stretching is important at leading order.
In a generic map the stretching is a strong function of space, so the
necessary time-rescaling would be position-dependent.

It is hard to see how the Lagrangian approach presented here could be used in
more realistic problems: perturbation theory was used extensively (which would
not be applicable in most real situations), both for computing the finite-time
Lyapunov exponents and the matrix exponential~\eqref{eq:Edef}.  We believe the
approach is instructive nonetheless, giving as it does a picture of the
strange eigenmode in Lagrangian coordinates.

Our approach does not yield much information about the long-time behavior of
the decay.  There is currently a debate as to whether the mechanism presented
in Refs.~\cite{Antonsen1996,Balkovsky1999,Son1999} gives a lower bound on the
decay rate~\cite{OttPrivate,FeredayPreprint}.  Our perturbation expansion
breaks down before this question can be answered.

\begin{acknowledgments}
The author wishes to thank Steve Childress for stimulating discussions.
\end{acknowledgments}


\appendix

\section{The Two-mode Solution}
\label{sec:twomode}

Though we have not found a general method of solution of~\eqref{eq:ADmap} with
the exponential given by~\eqref{eq:Edef}, there is at least an approximate
solution available that illustrates the broad features of a full solution.  It
also shows how the decay rate of the variance can become independent of the
diffusivity in the Lagrangian viewpoint, as in Ref.~\cite{Zeldovich1984}.

The method is based on defining a class of ``aligned'' wavenumbers (\ie, that
live inside the cone of safety), and retaining only two of these modes.  These
wavenumbers~$\kv^{(p)}$ are defined by
\begin{equation}
  k_\udir^{(p)} = \nugr^{p}\,\sin\theta\,,\qquad
  k_\sdir^{(p)} = \nugr^{-p}\,\cos\theta\,,
  \label{eq:kaction}
\end{equation}
that is,~$\kv^{(p)} = \ML^p\cdot\kv_0$, where~$\kv_0$ is any initial
wavenumber for large enough~$p$.  Then~$\kv$ satisfies
\begin{equation}
  \kv^{(p)} - \kv^{(p-1)} = \uvx\cdot\ML^{p-1},
\end{equation}
so that with the choice~$\kv=\kv^{(p)}$, $\lv=\kv^{(p-1)}$, the first
Kronecker delta in~\eqref{eq:Bdef} is unity for~$\iterj=p-1$.

At the~$\iter$th iteration, assume that only two wavenumbers are
important:~$\kv^{(\iter-d)}$ and~$\kv^{(\iter-d-1)}$.  The number~$d$ is a
``lag'' from the current iteration and will be adjusted later.

Define
\begin{equation}
  \mathcal{A}_d \ldef \exp\A^{(\iter)}_{\kv^{(\iter-d)}\kv^{(\iter-d)}}
  = \exp\l(-\scaledDiffc\l(\nugr^{2d}\,\cos^2\theta
  + \nugr^{-2d}\,\sin^2\theta\r)\r)
  \label{eq:cA}
\end{equation}
which is independent of~$\iter$, since we have defined~$p$ relative to the
current iteration of the map.  It can be shown that then the coupling
from~$\kv^{(\iter-d-1)}$ to~$\kv^{(\iter-d)}$ takes the simple form
\begin{equation}
  \mathcal{E}_d \ldef \eps\E^{(\iter)}_{\kv^{(\iter-d)}\kv^{(\iter-d-1)}}
  = -\tfrac{1}{2}\eps\l(\mathcal{A}_d - \mathcal{A}_{d-1}\r).
  \label{eq:cE}
\end{equation}

To recapitulate: at the~$\iter$th iteration, the mode~$\kv^{(\iter-d)}$ is
mapped to itself with coupling amplitude~$\mathcal{A}_d$,
and~$\kv^{(\iter-d-1)}$ is mapped to~$\kv^{(\iter-d)}$ with
amplitude~$\mathcal{E}_d$.  It is easy to show that in this two-mode situation
the decay rate is determined by the magnitude of~$\mathcal{E}_d$.  All that
remains is to find~$d$. \comment{NB XIII p. 111.}

The ``lag'' $d$ is obtained by maximizing~$\mathcal{E}_d$ over~$d$; $d$ has to
be large enough that~$\nugr^{2d}$ overcomes the tiny diffusivity
in~\eqref{eq:cA}---so the two~$\mathcal{A}$ terms don't cancel
in~\eqref{eq:cE}---but not so large that~$\mathcal{A}_d$ is damped.  We are
thus justified in approximating~\hbox{$\mathcal{A}_d \simeq
\exp(-\scaledDiffc\,\nugr^{2d}\,\cos^2\theta)$} (the other term
in~\eqref{eq:cA} is smaller by a factor~$\nugr^{-4d}$, which is small even
for~$d=1$).  We then have
\begin{equation}
  \mathcal{E}_d = \tfrac{1}{2}\,\eps(\Upsilon - \Upsilon^{\nugr^2}),
  \qquad
  \Upsilon \ldef \mathcal{A}_{d-1}\,,
\end{equation}
since~\hbox{$\mathcal{A}_d = (\mathcal{A}_{d-1})^{\nugr^2}$}.  This is easily
extremized over~$\Upsilon$: the maximum~$\lvert\mathcal{E}_d\rvert$ is
achieved for~\hbox{$\Upsilon=\nugr^{-2/(\nugr^2-1)}\simeq 0.7198$}, for
which~$\lvert\mathcal{E}_d\rvert \simeq 0.3074\,\eps$.  The lag is then given
by solving for~$d$ in terms of the extremizing~$\Upsilon$,
\begin{equation}
  d = 1 + \frac{1}{2\log\nugr}
  \log\l(\frac{\log\Upsilon^{-1}}{\scaledDiffc\cos^2\theta}\r)
  \simeq 0.5902 + 0.5195\log\scaledDiffc^{-1},
\end{equation}
which scales logarithmically with the diffusivity.  Note that the decay rate
is now completely independent of the actual value of the diffusivity: the lag
adjusts itself to compensate, introducing a separation of scale between the
dominant wavenumber~$\kv^{(\iter-d)}$ and the largest wavenumber in the
system,~$\kv^{(\iter)}$.

The actual decay rate as~$\scaledDiffc\rightarrow0$ (from the Eulerian
solution in~\cite{Thiffeault2003d}) is~$0.5\,\eps$ for small~$\eps$, compared
to the two-mode Lagrangian solution~$0.3074\,\eps$.  Thus, most of the
important behavior is captured by the two-mode solution.  The two modes can
be seen in the spectrum of the strange eigenmode in Fig.~\ref{fig:spectrum}:
the dominant peak at~\hbox{$\lVert\kv\rVert/k_\udir^{(\iter-2)} = 1$}
is~$\kv^{(\iter-d)}$ ($d=2$ in this case), and the peak
at~\hbox{$\lVert\kv\rVert/k_\udir^{(\iter-3)} = \nugr^{-1} \simeq 0.3820$}
is~$\kv^{(\iter-d-1)}$.  The other peaks are modes that could be included to
get a more accurate expression for the decay rate.

The two-mode solution also nicely illustrates the idea of the cone of safety:
both modes are always inside it, and because the cone is shrinking by a
factor~$\nugr^{-1}$ at each iteration then the modes have to follow suit.  The
key difference with~\cite{Zeldovich1984} is that here the concentration in the
modes is \emph{mapped} from one cone to another at each iteration, and is not
part of the initial condition.

\end{document}